\documentclass[12pt]{article}
\usepackage{graphicx}

\def\Re{{\cal R \mskip-4mu \lower.1ex \hbox{\it e}\,}}
\def\Im{{\cal I \mskip-5mu \lower.1ex \hbox{\it m}\,}}
\def\ie{{\it i.e.}}
\def\eg{{\it e.g.}}

\def\etal{{\it et al.}}

\def\sub#1{_{\lower.25ex\hbox{$\scriptstyle#1$}}}
\def\tev{\,{\ifmmode\mathrm {TeV}\else TeV\fi}}
\def\gev{\,{\ifmmode\mathrm {GeV}\else GeV\fi}}
\def\mev{\,{\ifmmode\mathrm {MeV}\else MeV\fi}}
\def\mpl{\ifmmode M_{pl}\else $M_{pl}$\fi}
\def\to{\rightarrow}

\def\subw{_{\rm w}}
\def\mh{\ifmmode m\sbl H \else $m\sbl H$\fi}
\def\mch{\ifmmode m_{H^\pm} \else $m_{H^\pm}$\fi}
\def\mt{\ifmmode m_t\else $m_t$\fi}
\def\mc{\ifmmode m_c\else $m_c$\fi}
\def\mz{\ifmmode M_Z\else $M_Z$\fi}
\def\mw{\ifmmode M_W\else $M_W$\fi}
\def\mws{\ifmmode M_W^2 \else $M_W^2$\fi}
\def\mhs{\ifmmode m_H^2 \else $m_H^2$\fi}   
\def\mzs{\ifmmode M_Z^2 \else $M_Z^2$\fi}
\def\mts{\ifmmode m_t^2 \else $m_t^2$\fi}
\def\mcs{\ifmmode m_c^2 \else $m_c^2$\fi}
\def\mchs{\ifmmode m_{H^\pm}^2 \else $m_{H^\pm}^2$\fi}
\def\ztwo{\ifmmode Z_2\else $Z_2$\fi}
\def\zone{\ifmmode Z_1\else $Z_1$\fi}
\def\mtwo{\ifmmode M_2\else $M_2$\fi}
\def\mone{\ifmmode M_1\else $M_1$\fi}
\def\tb{\ifmmode \tan\beta \else $\tan\beta$\fi}
\def\xw{\ifmmode x\subw\else $x\subw$\fi}
\def\ch{\ifmmode H^\pm \else $H^\pm$\fi}
\def\lum{\ifmmode {\cal L}\else ${\cal L}$\fi}
\def\inpb{\,{\ifmmode {\mathrm {pb}}^{-1}\else ${\mathrm {pb}}^{-1}$\fi}}
\def\infb{\,{\ifmmode {\mathrm {fb}}^{-1}\else ${\mathrm {fb}}^{-1}$\fi}}
\def\epem{\ifmmode e^+e^-\else $e^+e^-$\fi}
\def\ppb{\ifmmode \bar pp\else $\bar pp$\fi}
\def\bsg{\ifmmode B\to X_s\gamma\else $B\to X_s\gamma$\fi}
\def\bsll{\ifmmode B\to X_s\ell^+\ell^-\else $B\to X_s\ell^+\ell^-$\fi}
\def\bstt{\ifmmode B\to X_s\tau^+\tau^-\else $B\to X_s\tau^+\tau^-$\fi}
\def\lamt{\ifmmode \tilde\lambda\else $\tilde\lambda$\fi}
\def\shat{\ifmmode \hat s\else $\hat s$\fi}
\def\that{\ifmmode \hat t\else $\hat t$\fi}
\def\uhat{\ifmmode \hat u\else $\hat u$\fi}

\newskip\zatskip \zatskip=0pt plus0pt minus0pt
\def\matth{\mathsurround=0pt}
\def\lsim{\mathrel{\mathpalette\atversim<}}

\def\atversim#1#2{\lower0.7ex\vbox{\baselineskip\zatskip\lineskip\zatskip
  \lineskiplimit 0pt\ialign{$\matth#1\hfil##\hfil$\crcr#2\crcr\sim\crcr}}}

\def\grtsim{\,\,\rlap{\raise 3pt\hbox{$>$}}{\lower 3pt\hbox{$\sim$}}\,\,}
\def\lsim{\,\,\rlap{\raise 3pt\hbox{$<$}}{\lower 3pt\hbox{$\sim$}}\,\,}

\renewcommand{\thefootnote}{\fnsymbol{footnote}}

\begin{document} \begin{titlepage}
\rightline{\vbox{\halign{&#\hfil\cr
&SLAC-PUB-10510\cr
&July 2004\cr}}}
\begin{center}
\thispagestyle{empty} \flushbottom \centerline{ {\Large\bf Monte Carlo 
Exploration of Warped Higgsless Models 
\footnote{Work supported in part
by the Department of Energy, Contract DE-AC03-76SF00515}
\footnote{e-mails:
$^a$hewett@slac.stanford.edu, $^b$lillieb@slac.stanford.edu, and
$^c$rizzo@slac.stanford.edu}}}
\medskip
\end{center}

\centerline{J.L. Hewett$^{a}$, B. Lillie$^{b}$, and T.G. Rizzo$^{c}$}
\vspace{8pt} 
\centerline{\it Stanford Linear
Accelerator Center, 2575 Sand Hill Rd., Menlo Park, CA, 94025}

\vspace*{0.3cm}

\begin{abstract}
We have performed a detailed Monte Carlo exploration of the parameter
space for a warped Higgsless model of electroweak symmetry breaking in 5
dimensions.  This model is based on the $SU(2)_L\times SU(2)_R\times
U(1)_{B-L}$ gauge group in an AdS$_5$ bulk with arbitrary gauge kinetic
terms on both the Planck and TeV branes.  Constraints arising from
precision electroweak measurements and collider data are found to be
relatively easy to satisfy.  We show, however, that the additional
requirement of perturbative unitarity up to the cut-off, $\simeq 10$ TeV,
in $W_L^+W_L^-$ elastic scattering in the absence of dangerous tachyons
eliminates all models. If successful models of this class exist, they must 
be highly fine-tuned.
\end{abstract}



\renewcommand{\thefootnote}{\arabic{footnote}} \end{titlepage} 

%
%
%
%

\section{Introduction}

As the time of the LHC turn-on draws nearer, the search for
alternative theories to the standard single Higgs boson 
picture of electroweak symmetry breaking is intensifying.  One
such scenario \cite{warped} is particularly appealing in that it
employs a minimal particle content in a 5-dimensional spacetime
and exploits the geometry of the additional dimension to break
the electroweak symmetry.  The model is based on the
Randall-Sundrum framework \cite{RS} with an
$SU(2)_L\times SU(2)_R\times U(1)_{B-L}$ gauge group in
5-d Anti-de Sitter space.  The AdS$_5$ slice is bounded by
two branes, with the scale of physics on the IR(TeV)-brane being
given by $\Lambda_\pi\equiv \overline M_{Pl}e^{k\pi r_c}$, with
$k$ corresponding to the RS curvature parameter and $r_c$ being
the radius of the compactified dimension.
The set of boundary conditions, which
differ for the two branes, generate
the breaking chain $SU(2)_R\times U(1)_{B-L}\to
U(1)_Y$ at the Planck scale with the subsequent breaking
$SU(2)_L\times U(1)_Y\to U(1)_{QED}$ at the TeV scale.  The
electroweak symmetry is thus broken without the introduction
of a Higgs field. After the
Planck scale symmetry breaking occurs, a global $SU(2)_L\times
SU(2)_R$ symmetry is present in the brane description.  This
breaks on the TeV-brane to a diagonal group $SU(2)_D$ which
corresponds \cite{sundrum} to the custodial $SU(2)$ symmetry of the 
Standard Model (SM) and helps preserve the SM tree-level 
relation $\rho=1$. We denote this scenario as the
Warped Higgsless Model (WHM).   

In this scenario, the role of the Goldstone boson in generating
masses for the $W$ and $Z$ bosons is played by the would be 
zero-mode of the KK tower corresponding to the
$5^{th}$ component of the bulk gauge fields (\ie, $A^5_{0}$).
The $Z$ boson observed at LEP/SLC/Tevatron is the first
excitation of a neutral gauge boson KK tower, while the photon
corresponds to the massless zero-mode of this tower.
The $W$ boson observed in experiments is then the first state 
of a KK tower of charged gauge bosons, and there is no charged 
massless zero-mode.  The experimentally observed values
of the $W$ and $Z$ masses and couplings are essentially 
reproduced in this
model.  However, the presence of the gauge KK states affect a
number of processes.  In particular, much work has been performed
analyzing the contributions to the set of precision electroweak 
measurements in Higgsless scenarios 
\cite{NomuraI,Bar,DHLRI,NomuraII,CsakiTeV,BPRS,Chiv,nick,Casal}.  
In the flat space analog of the WHM \cite{flat}, 
\ie, a Higgsless model based on a flat higher 
dimensional spacetime, unacceptably large deviations from
precision electroweak data are generated.  However, good
agreement with the data can be obtained at tree-level in the
warped Higgsless scenarios, provided that
the masses of the KK excitations are sufficiently heavy. 
In addition, the KK excitations must satisfy the
constraints from direct production of new gauge bosons
at the Tevatron and from their contribution to contact 
interactions in four fermion processes at LEPII.

Note that precision observables are sensitive to one-loop electroweak
effects. In general, the loop corrections in this model will be qualitatively 
similar to
those in the SM (up to small shifts in the couplings). However, there are three
types of loop corrections that may cause large deviations: the gauge KK
excitations, the absence of loops with a physical higgs, and the top quark.
Since the gauge KKs are playing the role of the physical Higgs in WW
scattering, it is expected that they will do the same here, so the first two
effects should largely cancel. In our model all fermions are localized to the
Planck brane, and the parameters of the model are adjusted so the couplings are
as close to the SM couplings as possible. Hence, the top corrections should
again approximate the Standard Model values. (In a model where the top mass is
generated on the TeV brane {\cite {fermion}} a more careful treatment would
be needed.) Since we expect all loop corrections to qualitatively 
approximate the SM corrections, we will require that the tree level 
WHM approximate the tree level SM as closely as possible in the analysis 
below.

In the absence of a Higgs boson, or any other new physics, 
perturbative unitarity (PU) in elastic $W_L^+W_L^-$ scattering 
is violated at an energy scale of $\simeq 1.7$ TeV.  However, 
in these Higgsless scenarios, it is in principle possible
that the exchange of the neutral gauge
KK tower in $W_L^+W_L^-\to W_L^+W_L^-$ can restore PU, provided 
that a set of conditions on the KK masses and couplings are satisfied
\cite{flat}.  This works well in the flat space analog of the WHM,
but is problematic within the warped scenario.  In particular,
the region of parameter space which enjoys good agreement
with the precision electroweak and collider data leads to
low-scale ($\sim 2$ TeV) perturbative unitarity violation
(PUV) in gauge boson scattering \cite{DHLRI}. One
would at least expect the theory to remain perturbative up to the
cutoff scale of the effective theory on the TeV-brane, 
$\Lambda_\pi$,
where $\Lambda_\pi$ is roughly on the order of 10 TeV.  This leads to a
tension in the model parameter space in terms of finding
a region which simultaneously satisfies all of the model
building requirements as well as the experimental constraints.

In \cite{CsakiTeV,DHLRII} the effects of including 
the IR-brane terms
associated with the $U(1)_{B-L}$ and $SU(2)_D$ gauge
symmetries were examined; the presence of such terms
is known to alter the corresponding KK spectrum and couplings
\cite{DHRbt}.  In these analyses it was found that the addition
of the $U(1)_{B-L}$ IR-brane term could lead to improved agreement
with the electroweak data \cite{CsakiTeV} and the inclusion of
the $SU(2)_D$ brane term could delay PUV in $W_L^+W_L^-
\to W_L^+W_L^-$ to scattering energies of order $\sim 6-7$ TeV
\cite{DHLRII}.  However, a scan of the full parameter
space was not performed in order to determine whether there
exists a region where all the constraints discussed above
are simultaneously satisfied.

In this paper, we perform a detailed exploration of the WHM
parameter space via Monte Carlo techniques.  There are a
number of parameters present in this scenario:  (i) the set of
coupling strengths for each gauge symmetry: $g_{5L}$ which
is fixed by $G_F$, the ratio $\lambda\equiv g_{5B}/g_{5L}$
which is fixed by the value of $M_Z$, and the ratio
$\kappa\equiv g_{5R}/g_{5L}$ which lies in the restricted
range $0.75\lsim\kappa\lsim 4.0$, but is otherwise free.
(ii) The brane kinetic terms associated with the IR-brane,
$\delta_{B,D}$, and the UV-brane, $\delta_{L,Y}$. Here the brane 
terms will be treated as free phenomenological parameters but should 
in principle be calculable from the full theory once the UV-completion 
is known. The parameter
space is sufficiently large such that it is best scanned by
Monte Carlo sampling.  For each set of parameters, we subject
the model to a succession of tests: (i) model requirements,
such as the absence of ghosts and tachyon states, (ii) consistency
with the precision electroweak data, (iii) consistency with
the direct and indirect
collider bounds on new gauge boson production, and (iv) PU
in elastic $W_L^+W_L^-$ scattering.  In particular, we require
that this scattering process be unitary up to 
$\Lambda_\pi \simeq 10$ TeV. We find that the conditions
(i-iii) are relatively easy to simultaneously satisfy, but that
none of the models satisfied perturbative unitarity beyond the
scale of $\simeq 2$ TeV.   We conclude that if a successful
model of this type exists, it must be highly fine-tuned, or must contain 
other sources of new physics.

We present our analysis in the next two sections.
The formalism of the WHM is presented in detail in our earlier 
work \cite{DHLRI,DHLRII} and will not be reproduced here.

\section{Analysis: Electroweak and Collider Constraints}

As discussed above, the model in its present form contains five free 
parameters: $\kappa=g_R/g_L$, the ratio of the two $SU(2)_{L,R}$ 
gauge couplings which is 
expected to be of order unity, and the four brane kinetic term parameters, 
$\delta_{B,D,L,Y}$, corresponding to the various unbroken gauge groups on the 
TeV and Planck branes: $U(1)_{B-L}, SU(2)_D, SU(2)_L$ and $U(1)_Y$, 
respectively. Our approach is to choose a value for $\kappa$ and then 
explore the parameter space spanned by $\delta_{B,D,L,Y}$ via Monte Carlo 
methods. To be definitive we will assume that all the $\delta_i$ are 
constrained to lie in the range  
$-\pi kr_c \leq \delta_i \leq \pi kr_c$ as suggested in {\cite {DHRbt}}, 
and we fix $kr_c=11.27$ in our numerical 
study. For each set of values of the $\delta_i$ we define a successful 
model as one which passes through a number of cuts and 
filters that we now describe in some detail. Our results are compiled in 
Tables 1 and 2, 
which displays the amount of statistics generated for each value of 
$\kappa$ and the number of models which survive each successive constraint. 
Our statistics are concentrated near $\kappa=1$ as in this case the KK 
spectrum is relatively light and we are more hopeful that PU constraints 
will be satisfied. 

Upon generating a set of $\delta_i$ we first calculate a group of 
model parameters which are 
associated with the lightest massive charged and neutral gauge bosons, 
and ensure that they are  
identified with the observed SM fields, $W_1=W$ and $Z_1=Z$. We take the 
experimental values of their masses  
$M_Z=91.1875$ GeV and $M_W=80.426$ GeV {\cite {EWK}} as input to our analysis. 
This numerically fixes the low energy scale $k e^{-\pi kr_c}$ 
that gives the masses of the KK excitations 
in the RS model, as well as the value 
of the on-shell weak  mixing angle, $\sin^2 \theta_{OS}=1-M_W^2/M_Z^2$. 
Next, the requirement of the absence of ghosts in the unitary gauge of  
any physical theory implies that these two states, $W_1$ and $Z_1$, 
must have positive norms. Similarly, the field that represents the 
photon must also  have a positive norm. 
In addition to these constraints, we demand that  
the ratio of the squares of the gauge couplings, $\lambda^2=g_{B-L}^2/g_L^2$, 
be positive definite. As can be seen from Table 1, these few simple  
cuts can remove as much as $\sim 40\%$ of the parameter space volume.

Assuming that 
the SM fermions (at least the first two generations) are localized near the 
Planck brane we can now calculate a number of electroweak quantities. Recall 
that our philosophy is that we want the tree level Higgsless model to match 
the tree level SM as closely as possible, outside of the Higgs sector, 
since in 
many cases we expect approximately similar one-loop radiative corrections. 
As we found in our earlier works {\cite {DHLRI,DHLRII}}, 
a description of the $\gamma, W$ and $Z$ couplings to the 
SM matter fields can be parameterized in terms of two 
other definitions of the 
weak mixing angle in addition to $\sin^2 \theta_{OS}$. These two 
additional angles are defined via: $\sin^2\theta_{eg}=e^2/g_W^2$, with 
$g_W=g_{ffW_1}$ being the coupling of the $W$ to SM fermions on the 
Planck brane, and $\sin^2 \theta_{eff}$ being given by the couplings of the 
$Z$ to the same fermions at the $Z$-pole as discussed below. 
The electromagnetic coupling is, as usual,  
defined through the interaction of the massless neutral mode, $Z_0$, 
which we identify as the photon, to the SM fields. All three 
definitions of the weak mixing 
angle are identical at the tree level in the SM but are, in general, 
quite different numerically in the WHM.

\begin{table}
\begin{tabular}{l|rrrrrrrr}
\hline
Cuts\hspace{1.5cm}$\kappa$ & 0.75 & 0.9 & 1.0 & 1.25 & 1.5 & 2.0 & 3.0 
& 4.0\\
\hline
Initial Sample & 308,710 & 141,950 & 1,307,463 & 251,970 & 271,570 & 145,570
& 181,274 & 136,920\\
$\lambda^2 > 0$ & 130,286 & 62,202 & 585,011 & 115,455 & 125,035 & 67,662
& 82,583 & 16,204\\
No $\gamma,Z$ ghosts & 130,286 & 62,202 & 585,011 & 115,455 & 125,035 
& 67,662 & 82,583 
& 16,204\\
$|\delta\rho| < 0.005$ & 16,181 & 7,887 & 76,994 & 16,728 & 20,183 &
13,799 & 24,223 & 2,958\\
$|s^2_{\rm eff} - s^2_{\rm os}| < 0.005s^2_{\rm os}$ 
& 676 & 387 & 3,665 & 875 & 1,356 & 1,328 & 3,838 & 2,899\\
$|s^2_{\rm eg} - s^2_{\rm os}| < 0.005s^2_{\rm os}$ 
& 242 & 159 & 1,539 & 332 & 545 & 576 & 1,805 & 2,013\\
No $Z'$ ghost & 242 & 159 & 1,539 & 322& 545 & 576 & 1,805 & 2,013\\
$Z'$ Tevatron  & 150 & 102 & 1134 & 217 & 393 & 439 & 1,556 & 1,830\\
$m_{Z'} < 1.5$ TeV & 74 & 50 & 644 & 90 & 180 & 202 & 828 & 1,581\\
LEPII indirect & 70 & 45 & 550 & 72 & 80 & 175 & 796 & 1,178\\
Isospin coupling & 24 & 13 & 112 & 12 & 8 & 12 & 65 & 0\\
No Tachyons & 0 & 0 & 0 & 0 & 0 & 0 & 7 & 0  \\
PUV & 0 & 0 & 0 & 0 &  0& 0 & 0& 0 \\
\end{tabular}
\caption{Data samples and their responses to the various constraints 
as described in 
the text. The values represent the number of cases surviving each of the 
cuts.}
\end{table}

Writing the $Z$-pole couplings to SM fermions as
\begin{equation}
{g_Z\over {c_{OS}}} (T_3 -\sin^2 \theta_{eff}~ Q)\,,
\end{equation}
in obvious notation, we also can 
define an auxiliary quantity, $\rho_{eff}^Z=g_Z^2/g_W^2$, which relates 
the strengths of the $W$ and $Z$ gauge boson 
interactions. We identify $g_Z/c_{OS}=g_{ffZ_1}$. In the SM at tree-level 
$\rho_{eff}^Z=1$, of course. We note that all of the 
electroweak observables at the $W,Z$ mass scale can now be described in terms 
of the three weak mixing angles, $M_Z$, and 
$\rho_{eff}^Z$ and we have no need to 
introduce any other parameterizations.{\footnote {It can be easily shown that 
there is a unique mapping of the above parameters, together with $G_F$ which 
now involves a KK sum, over to 
the $\epsilon_i$ of Altarelli and Barbieri{\cite {AB}}.}} 
It is clear that if we wish to reproduce the 
tree-level SM we must have all three values of $\sin^2 \theta$ be almost 
equal as well as require that $\rho_{eff}^Z$ be very close to unity. In our 
numerical study we will demand that the three definitions of 
$\sin^2 \theta$ all 
be equal within $0.5\%$ and additionally 
require $|\delta \rho_{eff}^Z|=|\rho_{eff}^Z-1|$ to be less than 0.005.
The magnitude of these constraints 
should be comparable to the size of the one-loop 
generated electroweak corrections. 
This set of constraints is extremely powerful for the full WHM parameter 
space, but is especially 
strong for low values of $\kappa$ as can be seen from Table 1; only a few 
percent of the original model parameter space remains after 
applying these cuts. Note that models with larger values of $\kappa$ are 
generally favored by these electroweak constraints. This is not unexpected; 
we saw in our earlier work that 
the SM limit is approached rapidly as the value 
of $\kappa$ is increased. The price one pays for this is a rapid increase 
in the masses of the KK states leading to an obvious failure in PU as 
discussed below. 

\begin{figure}[htbp]
\centerline{
\includegraphics[width=9.5cm,angle=-90]{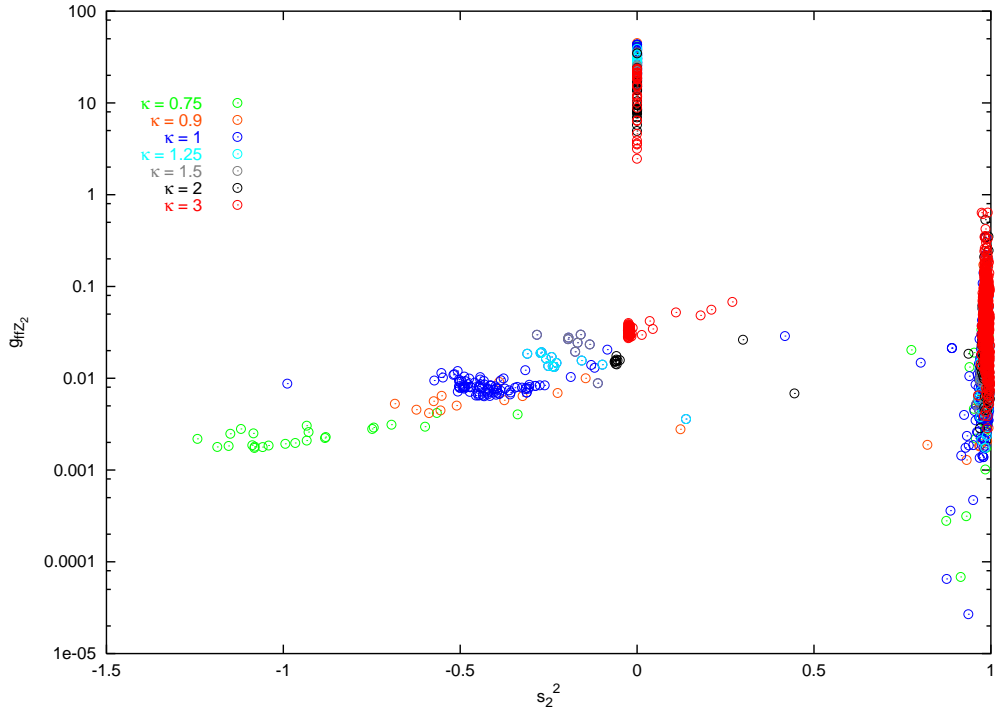}}
\vspace*{0.1cm}
\caption{The coupling strength of the first neutral KK excitation beyond the 
$Z$ in units of $g_W$ as 
a function of its effective $\sin^2 \theta$. The color 
coding labels models with different values of $\kappa$. All 
electroweak constraints have been applied to the cases shown as well as the  
bounds from the TeVII direct searches. A cut of $M_{Z_2}<1.5$ TeV has also 
been applied. Indirect LEPII constraints have not yet been imposed.}
\label{fig1}
\end{figure}

We now turn our attention to 
the mass and couplings of the next lightest neutral KK state, 
$Z_2$; these parameters are highly constrained by both experimental 
data as well as our requirement of PU as we will see below. 
We first impose the obvious constraints that this state not be a ghost and 
that it has not (yet) been observed in {\it direct} $Z'$-like production 
searches at LEPII or the Tevatron{\cite {TeV2,LEP2}}. 
This places a correlated cut on the mass of this state as a function of its 
couplings to the SM fermions on the Planck brane. 
Futhermore, we note that the {\it indirect} search constraints 
for $Z'$-like states must also be satisfied.   To be specific we will 
demand that the 
masses of the $Z_2$'s as well as their couplings to SM fermions are such 
as to have avoided the LEPII contact interaction constraints{\cite {LEP2}}.  
This constraint is actually quite powerful 
and removes an entire region in the $Z_2$ mass vs. coupling plane which 
survives the electroweak and Tevatron bounds. 
After imposing all these requirements we see from Table 1 that there are 
a respectable number of surviving cases.

\begin{figure}[htbp]
\centerline{
\includegraphics[width=9.5cm,angle=-90]{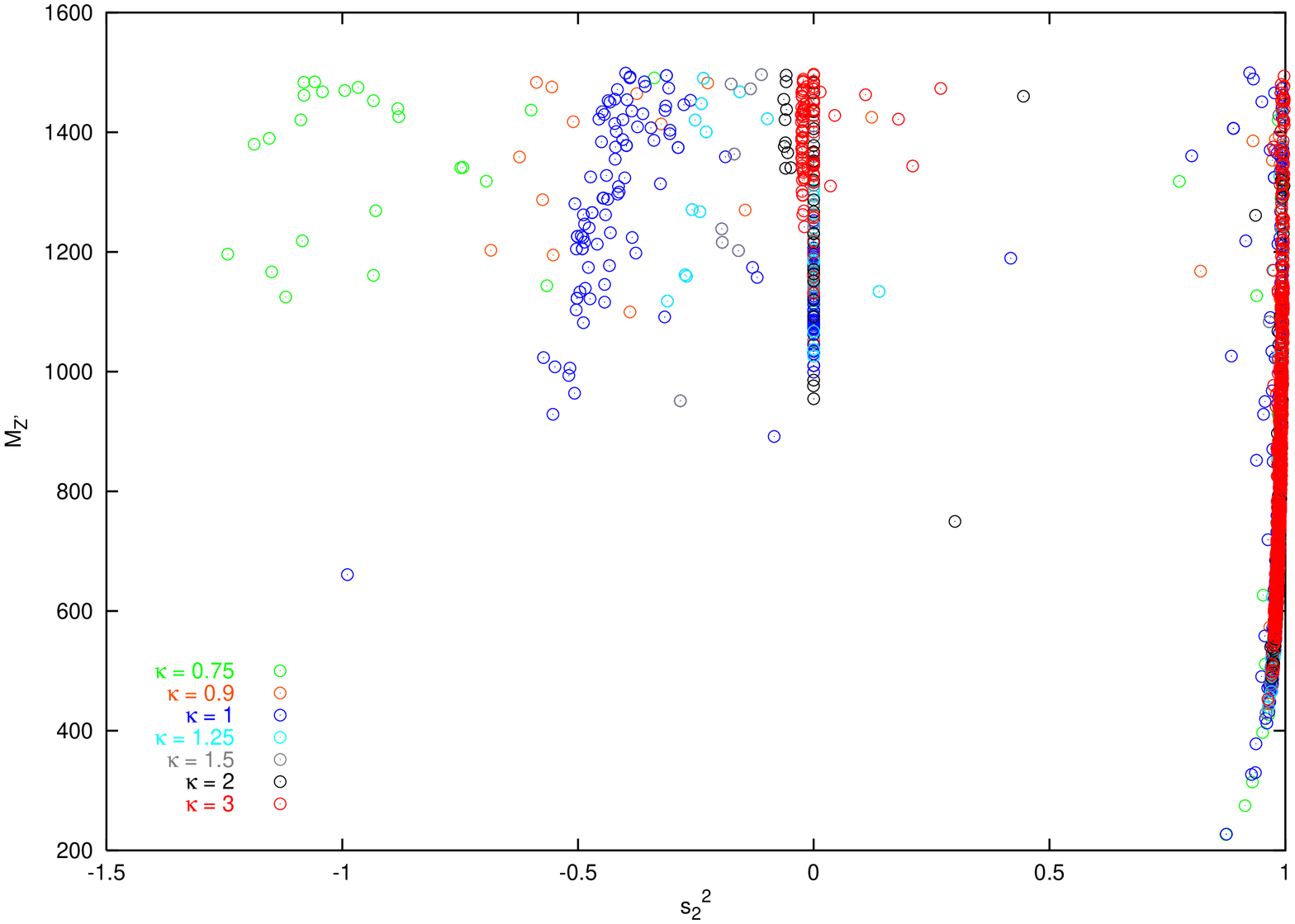}}
\vspace{0.4cm}
\centerline{
\includegraphics[width=9.5cm,angle=-90]{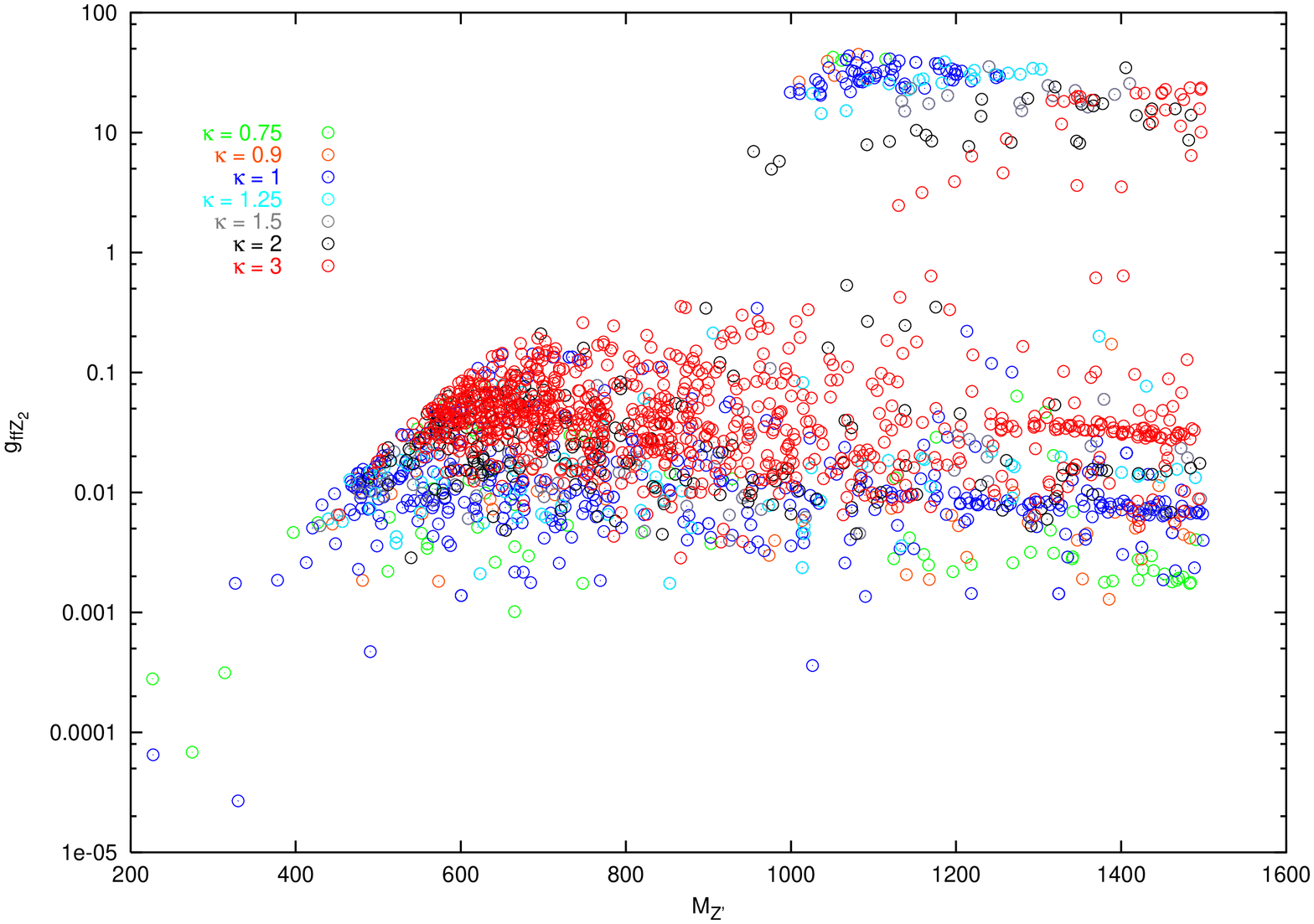}}
\vspace*{0.1cm}
\caption{Same as in the previous figure but now showing the mass-coupling 
strength and $s_2^2$ correlations.}
\label{fig2}
\end{figure}

The particular properties of the surviving cases will be examined in 
detail below. Figures. 1 and 2 
show the values of the $Z_2$ mass and coupling for those 
models passing all of our above cuts except the constraints imposed by 
LEPII; in these figures 
an additional requirement for PU that $M_{Z_2}\leq$ 1.5 TeV, to be 
discussed further below, has been imposed. The Tevatron direct search
constraint is responsible for the sharp diagonal boundary in the bottom
panel of Fig. 2.
Note that the couplings 
of the $Z_2$ can always be written in a form similar to the 
$Z$ above except we denote the overall coupling strength as 
$g_{ffZ_2}$ and the value of the corresponding weak mixing angle as $s_2^2=
\sin^2 \theta_{eff}(Z_2)$, \ie, $(g_{ffZ_2}/c_w)(T^f_{3L}-s^2_2Q^f)$. 
Note the large set of models near $s_2^2=0$ and 1, the former with large 
couplings and masses between $\sim$ 1 and 1.5 TeV. These strongly coupled 
cases are entirely removed by the LEPII contact interaction bound
and will not concern us further.  
It is important to note that at this point 
there {\it are} a reasonable of surviving models subsequent to  
applying this rather strict set of electroweak and collider 
constraints on the model parameter space. This situation is in contrast with 
results previously obtained by Barbieri \etal {\cite {BPRS}} in the case 
of the flat space analog where no warping is present. 
These authors showed that there was no significant 
region of parameter space which simultaneously satisfied the collider and 
precision electroweak constraints.  We have performed a Monte Carlo study 
of the flat space analog 
model, imposing the constraints presented above, and
effectively confirm their results. We note for completeness that if we  
strengthen our electroweak cuts such that $0.5\% \to 0.1\%$ in the analysis 
above, the number 
of surviving cases is reduced by a factor $\simeq 10$.

\section{Analysis: Perturbative Unitarity and Tachyons}

Unitarity is an important property of any gauge theory{\cite {bike,Hirn,Ohl}}. 
Before examining PU directly, two further filters can be applied that will 
help us to focus on models which may satisfy our basic requirements. 
If the $Z_2$ in any of the models that survive both the electroweak and 
collider constraints is to contribute significantly to the $W_L^+W_L^-$ 
amplitude it must predominantly couple to isospin and not to $B-L$ or
hypercharge $Y$.   Note that when $s_2^2$ 
is near unity(zero), the $Z_2$ couples almost purely to $Y$(isospin). 
To ensure that the $Z_2$ has significant isospin-like coupling,  
we will demand that $s_2^2 <0.7$.  We make exception for the  
special set of cases where the $Z_2$ mass is less than $\simeq 
400$ GeV. The reason for keeping these $B-L$-like coupled 
states is that their light masses 
may help induce a potentially large contribution to the $W_L^+W_L^-$ 
elastic scattering amplitude. Furthermore, there may exist somewhat heavier 
excitations not too far away in mass which {\it are} coupled to isospin. 
This cut on $s_2^2$ appears to be rather loose, but 
many of the surviving models have great difficulty satisfying it.
It is 
interesting to note that at this point in the parallel analysis of the flat 
space analog model {\cite {flat}} none of the cases satisfy this
constraint; all of the 
possible cases in the flat space model  
surviving both collider and electroweak constraints are found to essentially 
couple to $B-L$ or $Y$. 

In addition to the above, the $Z_2$ satisfying the collider constraints 
must still  be sufficiently 
light as to make a significant contribution to $W_L^+W_L^-$ elastic 
scattering. Recall that in the SM without a Higgs boson, PUV occurs near 
$\sqrt s \simeq 1.7$ TeV. This implies that there must be at least one, and 
more likely several, neutral KK states below this scale if they are to 
`substitute' for the SM Higgs in restoring unitarity. 
We thus impose the rather weak requirement 
that the lightest new neutral KK state, $Z_2$, must exist with 
a mass below 1.5 TeV; 
we make no further requirements on the spectrum for now. 

This pair of constraints on the mass and nature of the $Z_2$ 
couplings are rather 
difficult to satisfy simultaneously for the models that have passed 
the electroweak and collider 
cuts; relatively few cases survive at this point as can be seen from Table 1. 
Most of the models passing the electroweak and 
collider bounds tend to be either too massive 
or predominantly coupled to hypercharge. 
We can also see this from Figs. 1 and 2  
where 
the densely populated region near $s_2^2=1$ with a mass greater than 400 GeV 
is now removed by these cuts. At this point, the remaining models
are presented in Figs. 3 and 4; their distribution in $\delta_i$ space is 
shown in Fig. 5.
\begin{figure}[htbp]
\centerline{
\includegraphics[width=9.5cm,angle=-90]{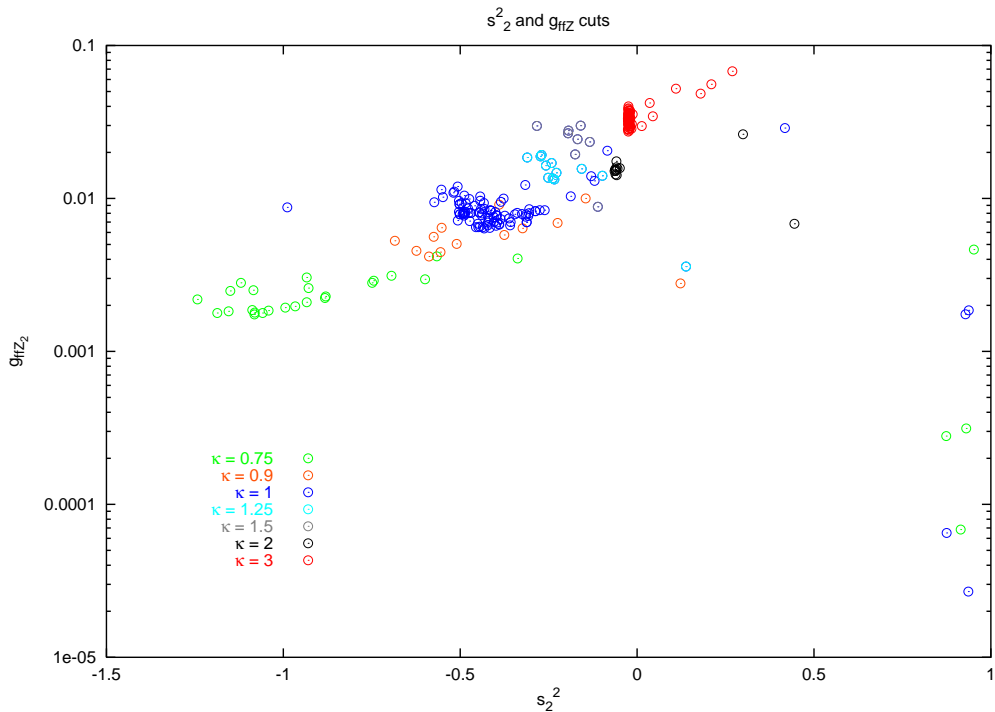}}
\vspace*{0.1cm}
\caption{Same as Fig.1, but now applying the constraints from LEPII and the 
correlated mass-$s_2^2$ cuts to remove the 
KK states coupling to $B-L$ and hypercharge as described 
in the text.}
\label{fig3}
\end{figure}

At this point we are 
ready to examine the PU characteristics of the remaining 
cases shown in Figs. 3-5 in detail.  
First we note that these models fall into two broad 
classes: those few with all 
positive $\delta_i$ and those with at least one of the $\delta_i$ being 
negative. An analysis of PU in $W_L^+W_L^-$ elastic scattering for the cases 
with all positive $\delta_i$ follows the standard procedure described in 
our earlier work {\cite {DHLRI,DHLRII}} which makes use of the 
scattering amplitude as given by {\cite {Duncan}} augmented to include 
additional neutral KK exchanges. Our proceedure is to calculate the complete 
amplitude using the expressions of Duncan \etal{\cite {Duncan}} which we 
modify to include an an arbitrary number of neutral KK exchanges in the $s-$ 
and $t-$ channels as well as an arbitrary $W$ 4-point coupling. We then 
integrate this amplitude to extract out the $J=0$ patial wave, $a_0$, subject 
to angular integration cuts imposed to avoid the photon $t-$channel pole. For 
our test of PU we demand that $|Re ~a_0|<1/2$, as is widely done in the 
literature. This analysis reveals 
that {\it none} of these models are much improved 
in comparison to the SM without a Higgs boson, \ie, PUV occurs at 
$\simeq 2$ TeV. The main reason for this is 
that these cases tend to have a light $Z_2$ which is predominantly coupled 
to $Y$ as discussed above. To restate, if the $\delta_i$ 
are all chosen positive, the models 
surviving the electroweak and collider constraints 
{\it do not} lead to theories which have PU beyond the $\simeq 2$ TeV scale.

\begin{table}
\begin{tabular}{l|rrrrrrrr}
\hline
Cuts\hspace{1.5cm}$\kappa$ & 1.0 & 1.5 & 2.0 & 2.5 & 2.75 & 3.0 & 4.0\\ \hline
Initial Sample & 611,150 & 304,680& 178,320 &122,801 
&266,801 & 169,862 & 70,661\\
$\lambda^2 > 0$ & 611,150 & 304,680 & 
178,320 &122,801 &266,801 &169,862 &70,661\\
No $\gamma,Z$ ghosts & 611,150 & 304,680 & 
178,320 &122,801 & 266,801 &169,862 &70,661\\
$|\delta\rho| < 0.005$ &168,732 &159,537 & 
107,709 &89,124 & 211,944 &146,087 & 69,867\\
$|s^2_{\rm eff} - s^2_{\rm os}| < 0.005s^2_{\rm os}$ 
& 0 & 502 & 1,506 & 2,734 & 8,600 & 7,456 & 8,724 & \\
$|s^2_{\rm eg} - s^2_{\rm os}| < 0.005s^2_{\rm os}$ 
& 0  &244 & 760 & 1,505 &4,887 & 4,308 &5,317\\
No $Z'$ ghost& 0  &244 & 760 & 1,505 &4,887 & 4,308 &5,317\\ 
$Z'$ Tevatron  &  0 & 6 &145 & 530 & 2,204 &2,233 &4,174 \\
$m_{Z'} < 1.5$ TeV & 0  & 6 &143 & 530 & 2,086 & 2,112 & 3,919\\
LEPII indirect & 0 & 6 & 143 & 509 &2,086  &2,112  & 3,919 \\
Isospin coupling & 0 & 0 &0 & 0 &0 & 0 & 0 \\
No Tachyons  & 0 &0  &0 & 0 & 0 & 0 &0  \\
PUV &0 &0 &0 &0 & 0& 0& 0\\
\end{tabular}
\caption{Same as the previous table but now for a special set of runs 
assuming all the $\delta_i \geq 0$. Note that many cases survive until the 
$B-L$ or $Y$ cut is employed.}
\end{table}

In order to verify this result we generated a larger statistical
sample,  an additional 
$\sim 1.7 \cdot 10^6$ models, distributed over various values of 
$\kappa$, assuming all of the $\delta_i \geq 0$. 
The results from performing these runs are shown in Table 2 using  
the same cuts as above. Here we see that although many models pass the 
combined 
collider and electroweak constraints none of them survive the non-$B-L/Y$ 
coupling requirement. Thus all of these models fail, confirming our 
previous results.  We checked that this  
also occurs in the analog flat space model. 

\begin{figure}[htbp]
\centerline{
\includegraphics[width=9.5cm,angle=-90]{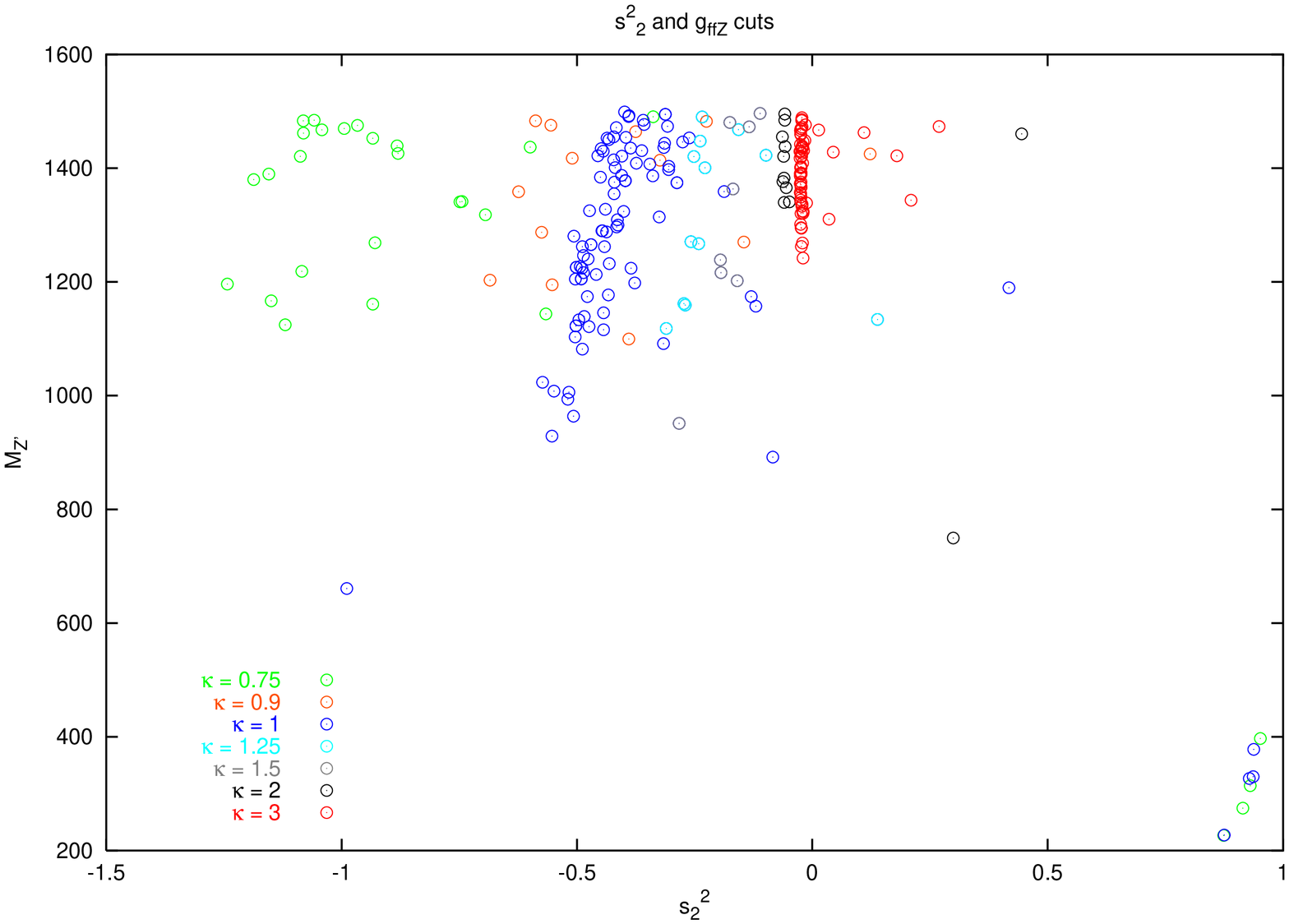}}
\vspace{0.4cm}
\centerline{
\includegraphics[width=9.5cm,angle=-90]{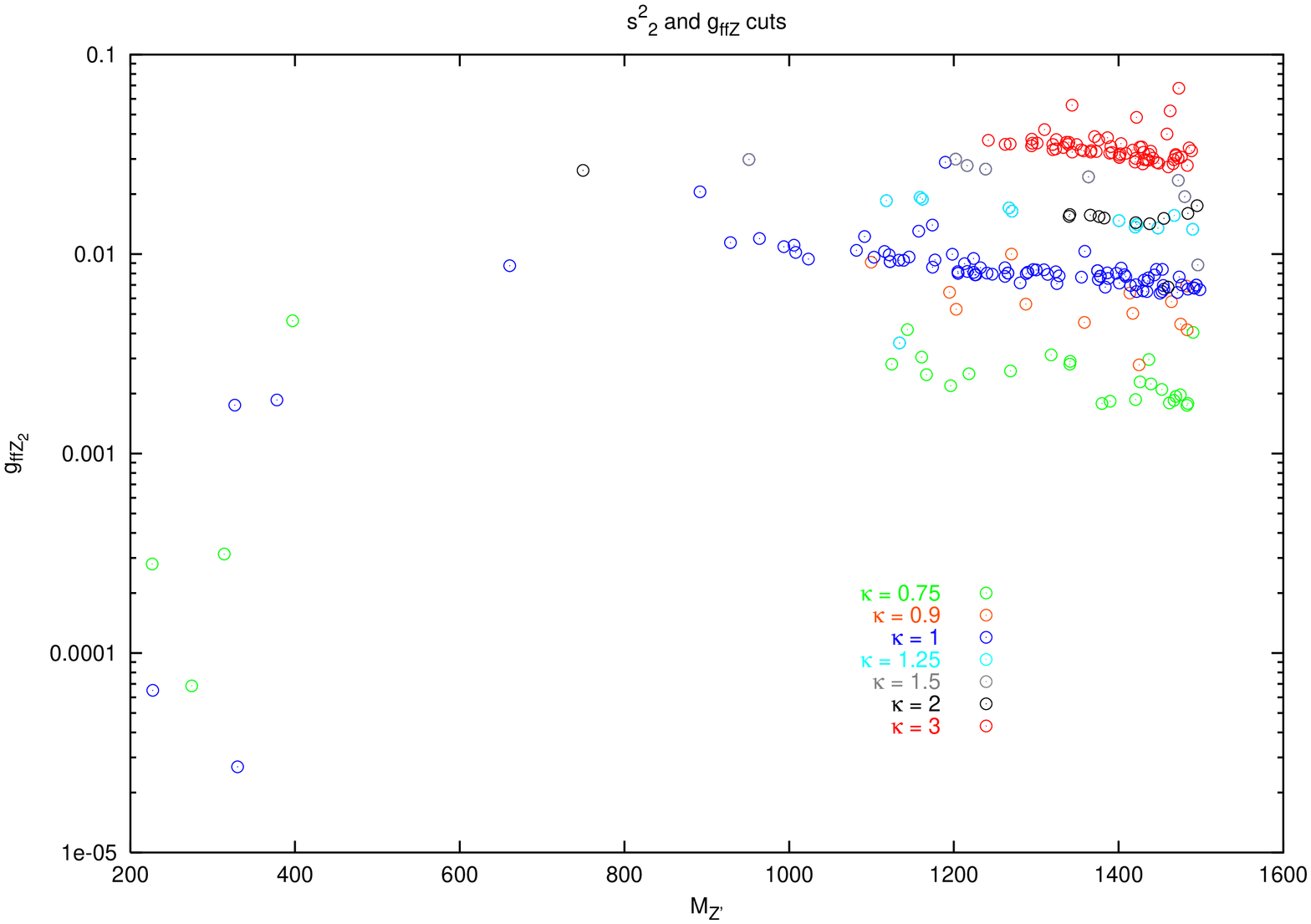}}
\vspace*{0.1cm}
\caption{Same as in the previous figure but now showing the alternate 
projections in the mass-coupling strength-$s_2^2$ parameter space.}
\label{fig4}
\end{figure}
\begin{figure}[htbp]
\centerline{
\includegraphics[width=9.5cm,angle=-90]{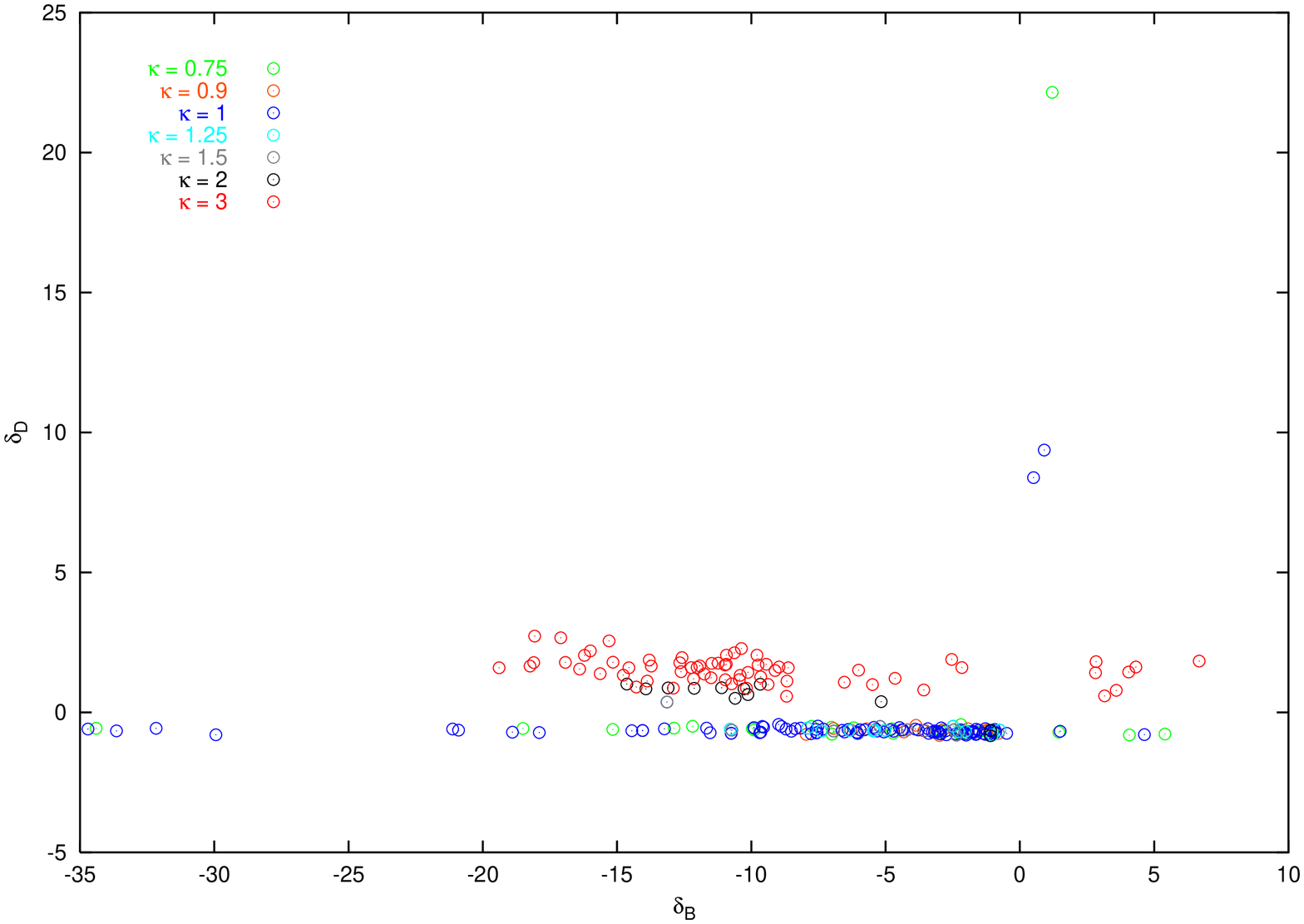}}
\vspace{0.4cm}
\centerline{
\includegraphics[width=9.5cm,angle=-90]{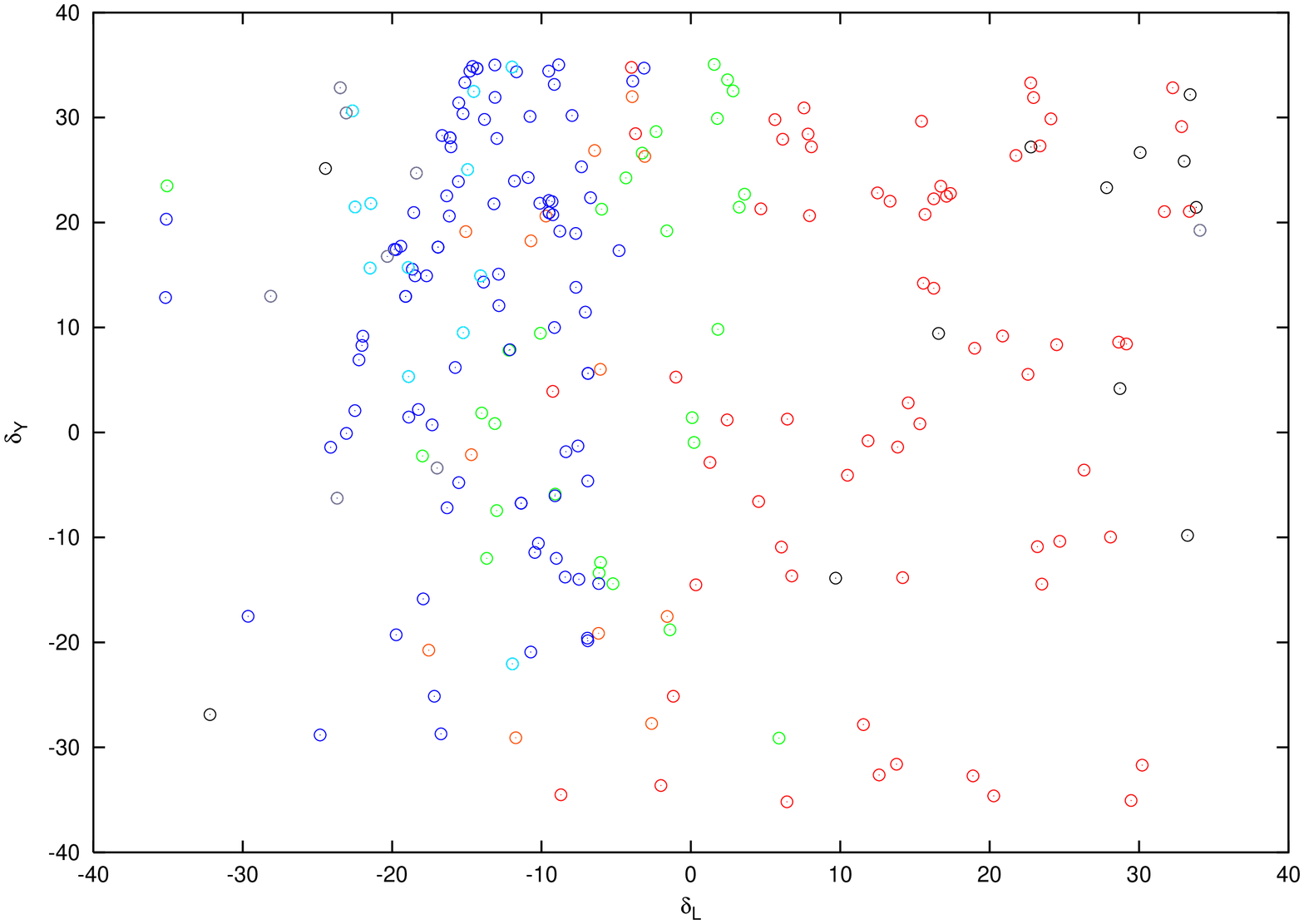}}
\vspace*{0.1cm}
\caption{Same as in the previous figure but now showing the values of  
the $\delta_i$ for the surviving models..}
\label{fig5}
\end{figure}

We now return to the models with at least one negative 
$\delta_i$.  Ordinarily, such models would not be considered since 
having negative $\delta_i$ at the tree-level implies the existence of 
tachyons with all their related difficulties {\cite {tachyon}}. 
Indeed, we have verified numerically that  
such tachyonic states do indeed exist in the spectrum for all the 
models in this class, and found that the tachyon 
masses are quite sensitive to the 
magnitudes of the $\delta_i$. Nomura {\cite {NomuraI}}, however, has 
argued that potentially large and negative boundary terms
associated with the Planck brane 
may be benignly generated at loop 
level without the significant influence of tachyons. These negative brane 
terms can be of sufficient 
importance numerically as to require their inclusion in a 
detailed tree-level analysis such as we are performing here. In such a case 
one could view 
the existence of tachyons as an artifact of including only partial one-loop 
effects. Since we are ignorant of the possible origin of negative 
$\delta_i$ in the full UV-completed theory, we must in principle consider 
these cases further. 

When analyzing the models with negative $\delta_i$, one has to take care 
that the existence of tachyons at the tree level does not have important 
phenomenological effects, \eg, tachyons that  
have significant couplings to SM fermions 
or which contribute substantially 
to  SM processes such as $W_L^+W_L^-$ scattering. At the 
very least, if we are to consider models with such states, the tachyons 
must be {\it truly} benign.   
Certainly models where these tachyons can lead to important physical effects 
must not be allowed. However, if the tachyons are significantly decoupled
from the SM fields we will consider such theories as benign and examine their 
PU properties.  
Based on the analysis of Nomura {\cite {NomuraI}},  
as well as our previous work {\cite {DHLRI,DHLRII}}, 
one might suspect that the tachyons induced by Planck brane kinetic terms,  
$\delta_{L,Y}$, are benign while those arising from kinetic terms 
on the TeV-brane, 
$\delta_{B,D}$, may not be. 

As an initial filter, we
analyze the couplings of tachyons to the SM fermions localized near the Planck 
brane;
clearly, these couplings can depend sensitively 
on the magnitude of the $\delta_i$. First, we must determine 
the number of tachyon states that are in the spectrum. 
In the flat space analog {\cite {flat}} 
of WHM it is easy to see that there can 
be only a single complex conjugate pair of tachyons 
in each of the neutral and charged KK towers; we 
expect this result to be equally valid in the warped case. A numerical study 
verifies these expectations and so we need to concern ourselves 
with only two tachyonic states, $T^0$ and $T^\pm$. 
We find that in all the sample cases examined the relevant 
couplings of these states to the SM fermions 
are suppressed by powers of $\epsilon=e^{-\pi kr_c}$ and are thus 
exponentially small. Such a result might have been 
anticipated since the Bessel functions of an 
imaginary argument, $I_n$ and $K_n$, which are 
needed to describe the tachyonic wavefunctions, are asymptotic to exponentials 
instead of sines and cosines as is case for the usual $J_n$ and $Y_n$. This  
suppression of couplings is similarly observed to take place in the flat 
space analog of the current model, though in a more modest 
fashion due to the absence of warping, 
where sinh's and cosh's replace the usual 
sines and cosines in the expression for the tachyonic wavefunctions. Thus 
consideration of the fermion-fermion-tachyon coupling places no additional 
constraints on any of the models under consideration. One should note, 
however, that such constraints may be of some importance in a wider class of 
models. 

\begin{figure}[htbp]
\centerline{
\includegraphics[width=9.5cm,angle=-90]{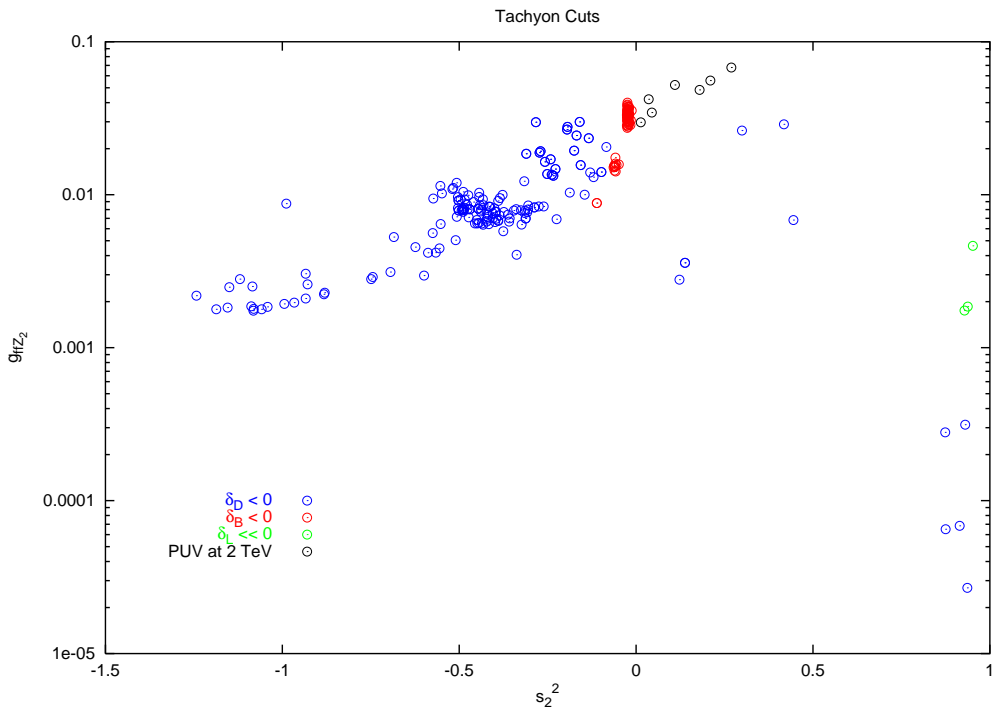}}
\vspace*{0.1cm}
\caption{Same as Fig.3 but now showing the effects of removing the models 
with potentially dangerous tachyons. The models surviving the tachyon cuts are 
also shown.}
\label{fig6}
\end{figure}

As a second test we turn to $W_L^+W_L^-$ elastic scattering. Here we expect 
a different result as the gauge fields are in the bulk and their wave 
functions sample the entire region between, as well as on, 
the two branes. A quick 
way to analyze this case is to consider the contribution of the neutral 
tachyon to the first sum rule of Csaki \etal\ {\cite {flat}}, which
is one of the conditions for PU. 
Their derivation of this sum rule relies heavily 
on the completeness of the set of eigenstates of Hermitian 
operators; thus the neutral KK spectrum in the case of $\delta_i <0$ is not 
complete unless the tachyon state is included. Clearly as the magnitude of 
the negative brane terms increase the couplings of the tachyon to SM gauge 
fields will also increase and the tachyon will become lighter. 
The important issue for us 
is whether or not the tachyon state makes a numerically 
{\it significant} contribution to the sum rule. 

Our results show that there are essentially three possibilities: ($i$) When 
$\delta_D <0$, we find that the tachyon makes a substantial contribution to 
the sum rule, which is on the order of $10\%$ or more 
of that of the photon, even when the magnitude of $\delta_D$ is small. 
In addition, this 
contribution is {\it negative}, \ie, the tachyon is also a ghost state! 
Certainly all such cases must be excluded. This is a powerful constraint 
as many of the surviving models 
shown in Figs. 3-5 have negative values of  $\delta_D$ in 
the region near $\sim -0.7$. ($ii$) When $\delta_{L,Y}<0$, the 
tachyon is generally 
sufficiently decoupled as to make almost no significant contribution to 
the sum rule. Not only are the couplings weak but the masses tend to lie in 
the multi-TeV range. 
For example, when $\delta_L \simeq -35$, a very  extreme value, 
the tachyon coupling to $WW$ is found to be $g_T^2 \sim 10^{-6}$ which 
is only dangerous if the tachyon is light. For $\delta_L$ values 
of lesser magnitude the couplings are significantly  smaller. 
This is as expected since we showed in our earlier work{\cite {DHLRI}}  
that in, \eg, a model with $\delta_L \sim -7.5$, the sum rules were very well 
satisfied without including any tachyonic contributions. 
Thus we will retain all such models for further study. 
($iii$) The 
remaining case where $\delta_B<0$ is a bit more problematic. As we saw in 
earlier {\cite {DHLRII}}, 
$\delta_B \neq 0$ has little influence on $W_L^+W_L^-$ 
elastic scattering since it only modifies the spectrum and couplings of the  
neutral KK's which couple predominantly to $B-L/Y$. 
The tachyon $W_L^+W_L^-$ coupling is found to be generally 
intermediate in strength between that of the $\delta_D <0$ case and 
those for $\delta_{L,Y}<0$, unless the magnitude of $\delta_B$ is 
reasonably large $\simeq 10$. Our analysis of the surviving sample of models, 
however, indicates that the values of $\delta_B$ are indeed of this magnitude 
or larger. Correspondingly the masses of these states are also dangerously 
light implying that they can significantly contribute to SM processes.  
We thus drop these cases from further consideration below.

Summarizing, our 
numerical study confirms our expectations that the tachyons induced by 
negative TeV-brane kinetic terms are dangerous while those induced by the 
corresponding Planck brane terms are benign unless $\delta_L$ 
is very near its lower bound. 
Figures 6 and 7 show what little remains of our surviving models 
after we employ the 
requirement that $\delta_{B,D} \geq 0$. These 10 cases are mostly clustered 
(those with large negative $\delta_Y$) at high $Z_2$ masses in excess of 
1.3 TeV and have pure isospin-like couplings. Those with negative 
$\delta_L$  tend to have much lighter $Z_2$ masses, of order 
less than 400 GeV.  
Their rather small couplings to fermions place them outside the range 
accessible to the Tevatron. These are the cases with small masses and 
large $B-L/Y$-like couplings that have survived the $B-L/Y$ cut imposed above. 
Unfortunately, these models 
all  have values of $\delta_L \simeq -\pi kr_c$ and 
thus have light tachyons with potentially significant couplings $\sim 10^{-6}$ 
and are thus dropped from further consideration.  This leaves 
only the 7 cases with negative $\delta_Y$ for further examination.

\begin{figure}[htbp]
\centerline{
\includegraphics[width=9.5cm,angle=-90]{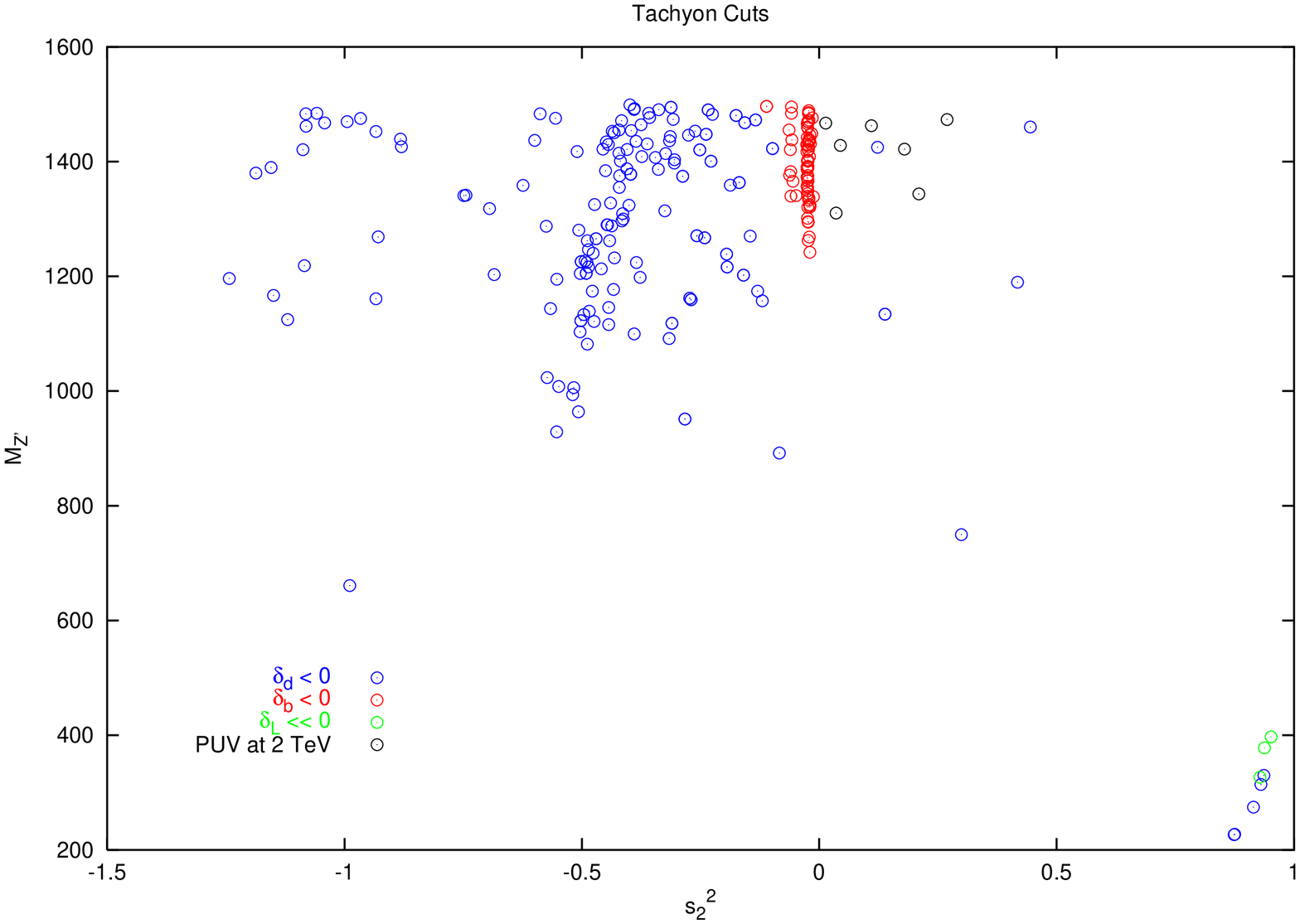}}
\vspace{0.4cm}
\centerline{
\includegraphics[width=9.5cm,angle=-90]{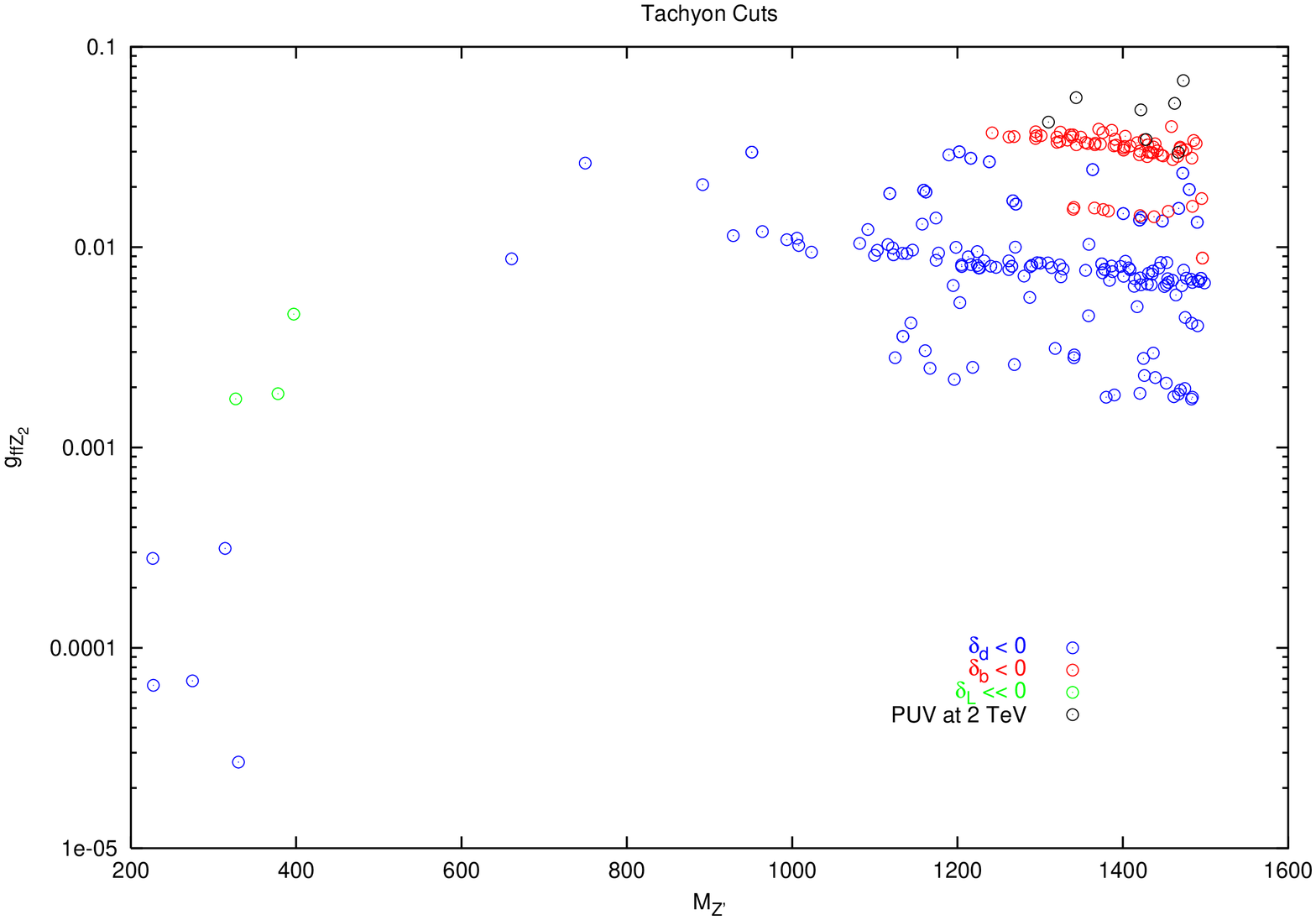}}
\vspace*{0.1cm}
\caption{Same as in the previous figure but now showing the alternate 
projections in the mass-coupling strength-$s_2^2$ parameter space.}
\label{fig7}
\end{figure}

We now turn to the PU characteristics of these few surviving models; 
naively we 
expect all these cases to be problematic since the first KK excitation is 
always in excess of 1.3 TeV even though they are isospin-like coupled. Indeed 
all of these cases 
lead to PUV in the 1.9-2.2 TeV range which is not a significant 
improvement over the case of the SM without a Higgs boson. We thus conclude 
that none of the surviving models pass our PU requirements leaving us with no 
remaining models. Note that in obtaining these results we have not required 
PU to be valid up to the cutoff but only that the successful model to 
reasonably better than the SM without a Higgs.

Since we found that PUV occured at $\sqrt s \simeq 2$ TeV in the surviving 
models it is interesting to compare this value to that of the cutoff scale, 
$\Lambda$, as determined by Naive Dimensional Analysis{\cite {pap}}, \ie, 
\begin{equation}
\Lambda=\epsilon {24\pi^3\over {g_{5L}^2}} \,,
\end{equation}
Following the notation of our earlier work, $g_W^2=N_\delta g_{5L}^2/2\pi r_c$,  
where $N_\delta$ is a number near unity which depends in detail on the 
values of the $\delta_i$. With $g_W$ fixed via $G_F$ this now yields 
\begin{equation}
\Lambda={12\pi^2\over {g_W^2}}  {N_\delta\over {kr_c}} {k\epsilon} \,,
\end{equation}
where $kr_c=11.27$ in our analysis. Taking typical model values we find that 
$\Lambda\simeq \Lambda_\pi \simeq 10$ TeV, which is much larger that the 
$\sqrt s$ values obtained above for PUV. Thus PUV is apparently lost long 
before the cutoff is reached.

\section{Conclusions}

In this paper we have performed a detailed tree-level Monte Carlo 
exploration of the 
parameter space of the 5-d warped Higgsless model which is based on the 
$SU(2)_L \times SU(2)_R \times U(1)_{B-L}$ gauge group in the Randall-Sundrum 
bulk.  We have generated several millions of test models allowing  
for arbitrary gauge kinetic terms on both the Planck and TeV branes which are 
parameterized through the $\delta_i$ coefficients. As we have seen from our 
earlier work this scenario suffers from a serious tension between 
constraints arising from precision electroweak measurements and collider 
data as well as the requirements of perturbative unitarity in $W_L^+W_L^-$ 
elastic scattering up to the $\Lambda_\pi \sim 10$ TeV scale. We have shown 
that it is relatively easy to find a class of models which satisfy all of 
the current direct and indirect collider bounds and 
yet has electroweak properties 
which are extremely close to those of the tree-level SM. 
As before, the size of the parameter space that satisfies the precision EW
constraints increases dramatically as $\kappa$ increases.
The real difficulty arises when we require the same 
theories to also satisfy perturbative unitarity 
while being free of dangerous tachyons. Though we have generated a fairly  
large data sample, none of the models we have examined have been able to 
satisfy all of our requirements simultaneously. 
We do note that if a generic solution of the PUV problem is found, there
appears to be enough room in the parameter space to accomodate precision EW
constraints.
Absent such a solution, we can thus conclude that 
either successful models of this type are highly fine-tuned or must
include additional sources of new physics  \cite{nandi} which
unitarizes the $W_L^+W_L^-$ scattering amplitude.

\noindent{\Large\bf Acknowledgements}

We would like to thank Csabi Csaki, Hooman Davoudiasl, Christophe Grojean, 
Tao Han, Graham Kribs, and John Terning for discussions related to this work. 
T. Rizzo would like to thank Atul Gurtu for his suggesting this type of 
analysis.

%
\def\MPL #1 #2 #3 {Mod. Phys. Lett. {\bf#1},\ #2 (#3)}
\def\NPB #1 #2 #3 {Nucl. Phys. {\bf#1},\ #2 (#3)}
\def\PLB #1 #2 #3 {Phys. Lett. {\bf#1},\ #2 (#3)}
\def\PR #1 #2 #3 {Phys. Rep. {\bf#1},\ #2 (#3)}
\def\PRD #1 #2 #3 {Phys. Rev. {\bf#1},\ #2 (#3)}
\def\PRL #1 #2 #3 {Phys. Rev. Lett. {\bf#1},\ #2 (#3)}
\def\RMP #1 #2 #3 {Rev. Mod. Phys. {\bf#1},\ #2 (#3)}
\def\NIM #1 #2 #3 {Nuc. Inst. Meth. {\bf#1},\ #2 (#3)}
\def\ZPC #1 #2 #3 {Z. Phys. {\bf#1},\ #2 (#3)}
\def\EJPC #1 #2 #3 {E. Phys. J. {\bf#1},\ #2 (#3)}
\def\IJMP #1 #2 #3 {Int. J. Mod. Phys. {\bf#1},\ #2 (#3)}
\def\JHEP #1 #2 #3 {J. High En. Phys. {\bf#1},\ #2 (#3)}


\begin{thebibliography}{99}
\bibitem{TeV2}
The strongest bounds at present are given by the D0 Collaboration in 
D0note 4375-Conf, v2.1 based on $\sim 200~pb^{-1}$ of Run II data.
\end{thebibliography}
\end{document}